\documentclass{emulateapj}

\usepackage{xspace}
\usepackage{graphicx}
\usepackage{rotating}
\usepackage{lscape}
\usepackage{natbib}
\usepackage{apjfonts}
\usepackage{color}

\def\errtwo#1#2#3{$#1^{+#2}_{-#3}$}

\newcommand\zth{$0^\mathrm{th}$}

\newcommand\chandra{\textsl{Chandra}\xspace}

\newcommand\integral{\textsl{INTEGRAL}\xspace}
\newcommand\isis{{\tt ISIS}\xspace}

\newcommand\mysou{IGR~J17511$-$3057}

\newcommand\rxte{\textsl{RXTE}\xspace}

\newcommand\swift{\textsl{Swift}\xspace}

\newcommand\xmm{\textsl{XMM-Newton}\xspace}

\newcommand\xspec{{\tt XSPEC}\xspace}
\newcommand\nthcomp{{\tt nthComp}\xspace}
\newcommand\bbodyrad{{\tt bbodyrad}\xspace}

\newcommand\diskline{{\tt diskline}\xspace}

\newcommand\aproxgt{\mathrel{%
     \rlap{\raise 0.511ex \hbox{$>$}}{\lower 0.511ex \hbox{$\sim$}}}}
\newcommand\aproxlt{\mathrel{%
     \rlap{\raise 0.511ex \hbox{$<$}}{\lower 0.511ex \hbox{$\sim$}}}}

\slugcomment{Accepted ApJ - June 2012 }
\shorttitle{A \chandra observation of IGR~J17511$-$3057}
\shortauthors{Paizis et al. 2012}

\begin{document}

\title{A \chandra observation of the millisecond X-ray pulsar \mysou }

\author{A. Paizis\altaffilmark{1}, 
M.~A. Nowak\altaffilmark{2}, 
J. Rodriguez\altaffilmark{3},
J. Wilms\altaffilmark{4},
S. Chaty\altaffilmark{3},
M. Del Santo\altaffilmark{5} and
P. Ubertini\altaffilmark{5}
}

\altaffiltext{1}{Istituto Nazionale di Astrofisica, INAF-IASF, Via Bassini 15,
20133 Milano, Italy; ada@iasf-milano.inaf.it}
\altaffiltext{2}{Massachusetts Institute of
  Technology, Kavli Institute for Astrophysics, Cambridge, MA 02139,
  USA; mnowak@space.mit.edu}
\altaffiltext{3}{AIM  - Astrophysique, Instrumentation et Mod\'elisation 
(UMR-E 9005 CEA/DSM-CNRS-Universit\'e Paris Diderot)
Irfu/Service d'Astrophysique, Centre de Saclay 
FR-91191 Gif-sur-Yvette Cedex, France}
\altaffiltext{4}{Dr.~Karl Remeis-Sternwarte and Erlangen Centre for
  Astroparticle Physics, Universit\"at Erlangen-N\"urnberg,
  Sternwartstr.~7, 96049 Bamberg, Germany}
\altaffiltext{5}{IAPS, INAF, Via Fosso del Cavaliere 100 00133 Rome Italy}

\begin{abstract}
 \mysou~is  a low mass X-ray binary hosting a neutron star and is one of the few  accreting millisecond X-ray pulsars with X-ray bursts. We report on a 20\,ksec \chandra~grating observation of \mysou, performed on 2009 September 22. 
We determine the most accurate  X-ray position of \mysou, \mbox{$\alpha_\mathrm{J2000}$=17$^\mathrm{h}$ 51$^\mathrm{m}$ 08$^\mathrm{s}$.66},
\mbox{$\delta_\mathrm{J2000}$= $-$30$^{\circ}$ 57$^{\prime}$ 41.0$^{\prime\prime}$}(90\% uncertainty  of 0.6$^{\prime \prime}$).
During the observation, a $\sim$54\,s long type-I X-ray burst is detected. The persistent (non-burst) emission has an absorbed 0.5--8\,keV luminosity of 1.7$\times$10$^{36}$ $\mathrm{erg~s^{-1}}$ (at 6.9\,kpc) and can be well described by a thermal Comptonization model of  
soft, $\sim$0.6\,keV, seed photons up-scattered by a hot corona. The type-I X-ray burst spectrum, with average luminosity over the 54\,sec duration $L_\mathrm{0.5-8\,keV}$=1.6$\times$10$^{37}$ $\mathrm{erg~s^{-1}}$, can be well described by a blackbody with $kT_\mathrm{bb}\sim$1.6\,keV and $R_\mathrm{bb}\sim$5\,km. While an evolution in temperature of the blackbody can be appreciated throughout the burst (average peak $kT_\mathrm{bb}$=\errtwo{2.5}{0.8}{0.4}\,keV to tail $kT_\mathrm{bb}$=\errtwo{1.3}{0.2}{0.1}\,keV), the relative emitting surface shows no evolution. \\
The overall persistent and type-I burst properties observed during the \chandra~observation are consistent with what  was previously reported during the 2009 outburst of \mysou.

\end{abstract}


\section{Introduction}\label{sec:intro}
\setcounter{footnote}{0}

Low Mass X-ray Binaries containing a neutron star (hereafter NS LMXBs) are very old systems (10$^{8}-10^{9}$\,yrs), with a NS magnetic field that is believed to have decayed to about 10$^{8}-10^{9}$\,G. It is believed that since the NS spends a substantial fraction of its life accreting gas via an accretion disk, it is finally spun-up to millisecond levels \citep{tauris06}. 
This belief is supported by the fact that in 23 cases \citep[out of more than 150 known LMXBs,][]{liu07} the NS spin frequency has been detected at the millisecond level \citep[see][for a complete list]{patruno10}. These detections support the scenario that  LMXBs are the progenitors of millisecond radio pulsars with a low magnetic field. 

When pulsations occur during surface thermonuclear explosions, known as type-I X-ray bursts 
\citep[see][for a review]{strohmayer06}, the sources are known as \emph{nuclear powered X-ray pulsars} (hereafter NPXP).
Up to now, pulsations during bursts
(called burst oscillations) have been detected in 15 sources \citep{altamirano10}.
During these events, the 
accumulated nuclear fuel first ignites at the point of
the neutron star surface where it reaches the critical ignition column density and then spreads
to all adjacent areas on the surface. When nuclear burning occurs uniformly over the surface, no
\lq\lq hot spot\rq\rq~is created and the neutron star spin will still be hidden. But in some cases a \lq\lq patchy\rq\rq~burning process can occur, making the neutron star spin period visible.  With the decrease
of the X-ray burst flux, the non-uniformity fades out and so do the pulsations. 

 In other cases, pulsations occur in the \lq \lq persistent\rq \rq~X-ray emission  (i.e., not during type-I X-ray bursts), and the sources are known as \emph{accreting millisecond X-ray pulsars} (hereafter AMXP). Up to now, 14 such sources have been detected \citep[all transient X-ray sources,][]{patruno10b,papitto11} and it is believed that matter from the accretion disk is channeled by the magnetic field lines onto the magnetic poles, forming a hot spot visible in X-rays. An important detection for our comprehension of the pulsating mechanism in AMXPs (versus the non pulsating majority of LMXBs) was achieved with the discovery of pulsations that were not detected  throughout the outburst, but only intermittently, e.g., as HETE~J1900.1$-$2455 \citep{kaaret06}, Aql~X$-$1 \citep{casella08} and SAX~J1748.9$-$2021 \citep{altamirano08}. These sources are important because they may be the intermediate link between persistent AMXPs and non pulsating neutron star LMXBs.

Of the currently known 14 AMXPs, only five, including \mysou, belong to both the NPXP and AMXP classes, i.e., show pulsations during type-I X-ray bursts {\it and} during the persistent (non-burst) emission \citep{altamirano10}.\\

On 2009 September 12 (MJD 55087) \integral~discovered a new hard X-ray source, \mysou~\citep{baldovin09}, 
detected during the \integral~Galactic bulge monitoring program \citep{kuulkers07}. 
Shortly thereafter, we reported  the best X-ray position of the source
 from a preliminary analysis of our \chandra~data:
\mbox{$\alpha_\mathrm{J2000}$=17$^\mathrm{h}$ 51$^\mathrm{m}$ 08$^\mathrm{s}$.66},
\mbox{$\delta_\mathrm{J2000}$= $-$30$^{\circ}$ 57$^{\prime}$ 41.0$^{\prime\prime}$} 
 \citep[90\% uncertainty of 0.6$^{\prime\prime}$,][]{nowak09}. Near infrared follow-up observations identified within the \chandra~error box the counterpart at a magnitude of $K_\mathrm{s}$=18.0$\pm$0.1 \citep{torres11a,torres11b}, but 
no radio counterpart was detected with a 3$\sigma$ upper limit of 0.10\,mJy \citep{miller09}. 

Shortly after the discovery, pulsations at 245\,Hz  were reported \cite[][using \rxte~data]{markwardt09}, as well as the first type-I X-ray burst \cite[][\swift data]{bozzo09}, and burst oscillations very close to the  neutron star spin frequency \cite[][\rxte~data]{watts09}, making \mysou~the fifth LMXB hosting a neutron star belonging to both the AMXP and NPXP classes. 

Similarly to other AMXPs, \mysou~can be classified as an atoll source based on its timing and spectral characteristics \citep{bozzo09,papitto10,kalamkar11,ibragimov11,falanga11}.

\begin{figure}
\epsscale{1.0} \plotone{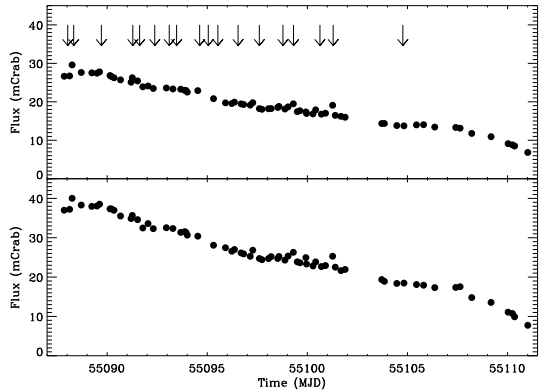}
\caption{Outburst of \mysou~as observed by \rxte: 2--10\,keV (upper panel) and 10--25\,keV 
(lower panel) intensity, normalized to the Crab. The vertical arrows indicate the times of detected type-I X-ray bursts \citep{falanga11}. The \chandra~data studied here were obtained on 2009 September 22 (MJD 55096).}
 \label{fig:lcr_all}
\end{figure}

The source faded beyond \rxte~detection limit after 2009 October 8 \citep[MJD 55113,][]{markwardt09}, with  coherent pulsations detected throughout all the outburst. Recently, possible twin kHz quasi-periodic oscillations (QPO) have been reported \citep{kalamkar11}. 
During the whole outburst, a total of 18 type-I X-ray bursts have been detected, marked as vertical arrows in Fig~\ref{fig:lcr_all}: ten by \rxte, three by \swift~(one of which in common with \rxte), two by \xmm, three by \integral, and one by \chandra~\citep[see][for a complete list]{falanga11}. With the exception of the \chandra~one, 
all the type-I X-ray bursts have been previously studied and reported  \citep{bozzo09,altamirano10,papitto10,falanga11,riggio11}. 

In this paper we focus on the unpublished \chandra/HETG observation of \mysou~(2009 September 22, MJD 55096) that we triggered as part of our approved \chandra~target of opportunity program.

\section{The data}\label{sec:data}

We observed \mysou~for 20\,ks with \chandra on 2009 September 22, from 07:40:39  UT until 13:32:03 UT
with the High Energy Transmission Grating Spectrometer, HETGS \citep{canizares00} collecting high resolution spectral information with the  High Energy Grating, HEG 0.8--10\,keV, 
and Medium Energy Grating, MEG 0.4--8.0\,keV.
The data were analyzed in a standard manner, using the CIAO version 4.3 software 
package and \chandra CALDB version 4.4.6. The spectra were analyzed with 
the \isis analysis system, version 1.6.1 \citep{houck02}. For  pileup correction in the presence of high fluxes (such as the type-I X-ray burst), we used the S-lang script {\sl simple\_gpile2},  within the \isis fitting package, as described in \cite{nowak08} and \cite{hanke09}. The \chandra \zth-order spectrum  was not used in the spectral analysis as it severely suffers from pile-up, especially in the burst phase. Given the source brightness and the intrinsic low \chandra~background, no background removal was applied.

To develop a feeling of the overall flux evolution of  \mysou, we reduced  
available \rxte~data \citep[same dataset of][]{altamirano10}.  Standard filtering criteria were applied \citep[see, e.g.,][]{rodriguez08} 
and the average count rate was obtained from the top layer of PCU2 for 
each individual observation. Light curves from the Crab 
nebula and pulsar from the two closest observations were also extracted and used  to renormalize 
the PCA source count rate to the Crab one. \\

\section{Results}\label{sec:results}

\subsection{The X-ray position of \mysou}\label{sec:position}
We extracted the X-ray position of \mysou~from the \zth-order image
obtaining  \mbox{$\alpha_\mathrm{J2000}$=17$^\mathrm{h}$ 51$^\mathrm{m}$ 08$^\mathrm{s}$.66},
\mbox{$\delta_\mathrm{J2000}$= $-$30$^{\circ}$ 57$^{\prime}$ 41.0$^{\prime\prime}$}, consistent with what we had 
reported in  \cite{nowak09}.

\begin{figure}
\epsscale{1} \plotone{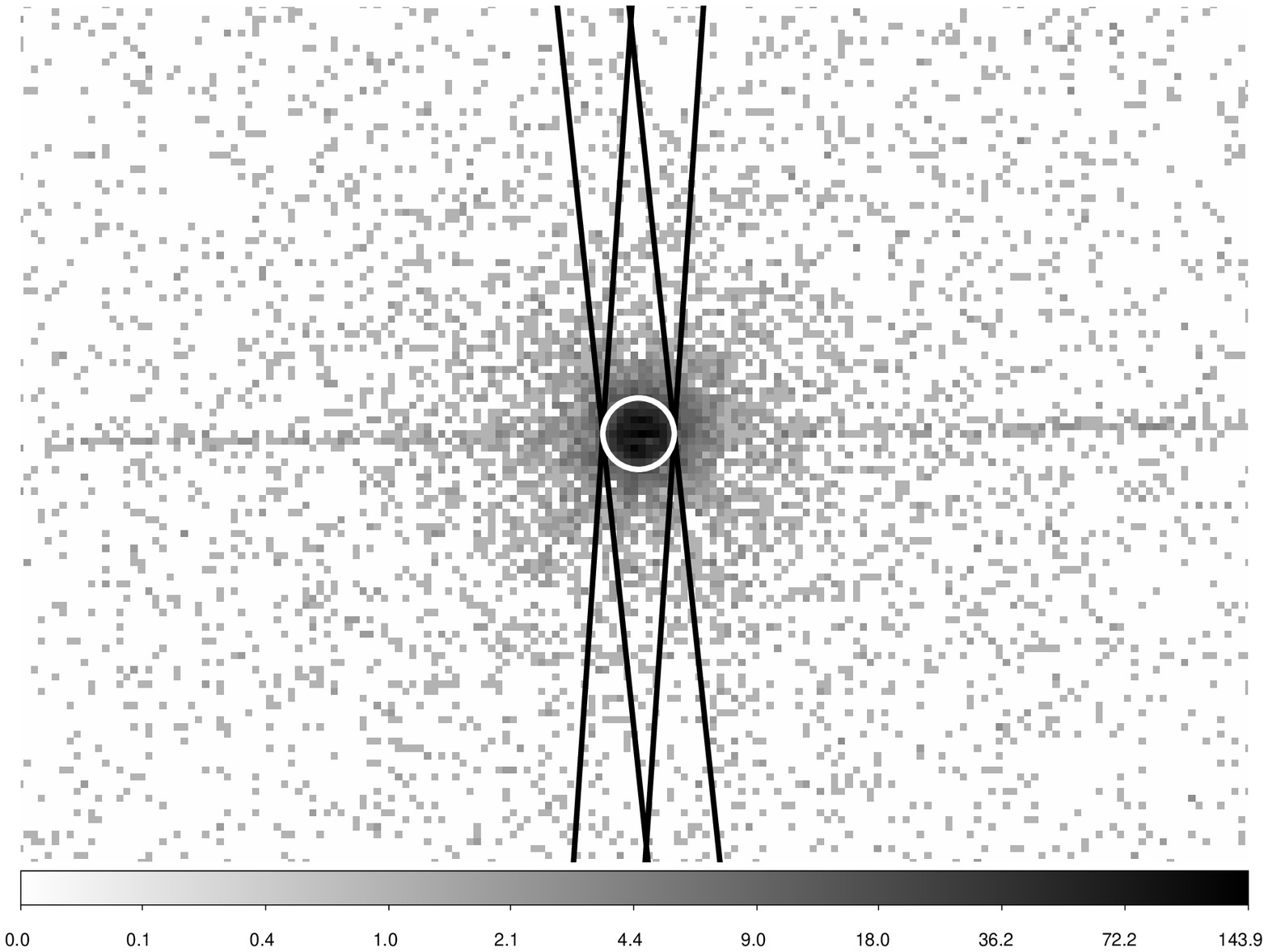}
\epsscale{1} \plotone{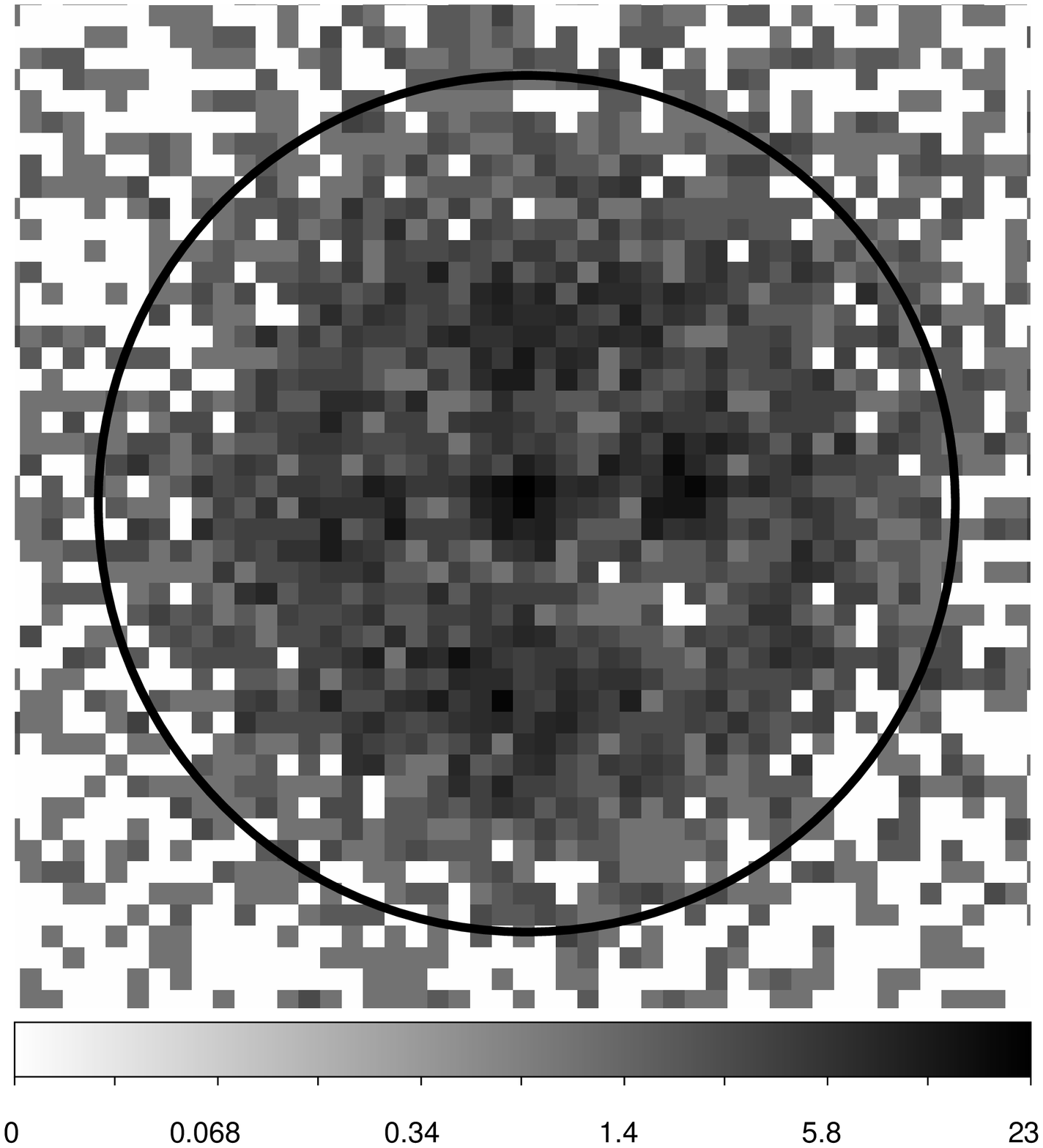}
\caption{\chandra~zeroth order images of \mysou. The locations
of the zeroth order and grating arms regions are shown. See text.}
 \label{fig:pos}
\end{figure}

Fig.~\ref{fig:pos} shows the zeroth order image at $0.5^{\prime\prime}$ (upper panel) and $0.12^{\prime\prime}$  (lower panel) binning (burst plus persistent emission). There is apparent structure in these images due to pileup, especially during the burst portion of the lightcurve. For this reason, the source location was determined by intersecting the readout streak (visible in the $0.5^{\prime\prime}$ image) with the grating arms (outside of the field of view covered in the image, but 
$5^{\prime\prime}$  wide boxes along the arm positions are shown). This was accomplished with the  \texttt{findzo} algorithm, which is standardly used for determining the zeroth order position when a readout streak is strongly detected (and hence pileup is affecting the zeroth order image). This is discussed more extensively in \cite{huenemoerder11}, who discuss its use
in the \chandra Transmission Gratings Catalog (TGCAT). Its estimated positional uncertainty is $<0.1^{\prime\prime}$. 
 We note that cross-correlation of Sloan Digital Sky Survey (SDSS) source positions with those obtained from the \chandra
Source Catalog (CSC) -- which relies upon \chandra absolute astrometry, as we use here -- require only a $0.16^{\prime\prime}$ 1-$\sigma$ systematic correction (i.e., $0.26^{\prime\prime}$ at 90\% confidence level) to bring the SDSS and CSC positions into statistical agreement. \citep[see Fig. 22 of][]{primini11}.

The error we obtain is, however,  significantly less than the $0.6^{\prime\prime}$ 90\% confidence level uncertainty
 claimed for \chandra absolute astrometry\footnote{http://cxc.harvard.edu/cal/ASPECT/celmon/}  when no other sources are present in the field of view for refined registration of the field, as is the case for this observation.
We, therefore, attribute to the position found a 90\% uncertainty of 0.6$^{\prime \prime}$.

\subsection{The type-I burst profile}\label{sec:theburst}
An overview of the outburst of \mysou~obtained from   \rxte~data 
is shown in Fig.~\ref{fig:lcr_all}. The \chandra~data presented here occurred on 2009 September 22 (MJD 55096 in the plot). 

\begin{figure}
\epsscale{1} \plotone{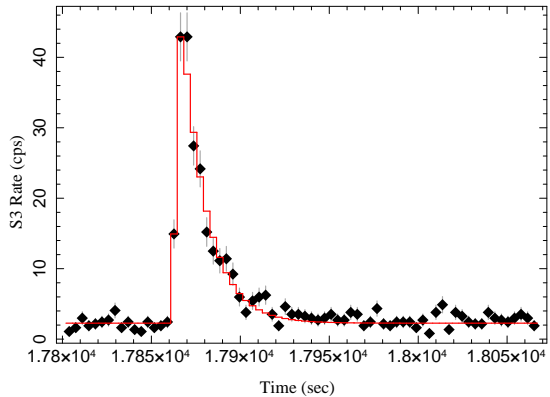}
\caption{Zoom on the type-I X-ray burst observed in our \chandra~observation of \mysou~(2--8\,keV). The time  is relative to the first photon arrival time of the observation.}
 \label{fig:burst}
\end{figure}

A zoom in the lightcurve of \mysou~as obtained during our \chandra~observation, including the type-I X-ray burst detected at 2009 September 22 12:54:56 UTC, and lasting for about 54\,s, is shown in Fig.~\ref{fig:burst}.  At the time of the burst the source was not being observed by \rxte.

Fitting the burst profile with a Fast Rise Exponential Decay (hereafter FRED) function, we obtain: a start time of the burst $t_\mathrm{start}$=(17861$\pm$1)\,s, relative to the first photon arrival time of the observation, which translates in 2009 September 22   12:54:56 UTC; a rise time of the burst, $t_\mathrm{rise}$, within the range of (3.6--6.3)\,s and a decay time of the burst $t_\mathrm{decay}$=(14$\pm$1)\,s.  

In order to minimise the offset of photon arrival times due to the fact that the CCD chips are read out quasi serially in the timed exposure (TE) mode, the above results on the FRED properties were obtained using the S3 chip alone. Rectangular regions along the arms (20$\times$90 pixel boxes) and a circular region of 30 pixels around the \zth-order (excising the innermost 16 pixel radius to avoid pile-up in the non-dispersed photon region) were used, together with a two-frame bin time, i.e., 3.68\,s  (1.84\,s$\times$2\,frames). 

Given the very sharp flux increase of the burst 
\citep[Fig.~\ref{fig:burst} and $t_\mathrm{rise}<$1.2\,s for other reported bursts from the source,][]{altamirano10}, it is clear that the binning time used for the fit is likely to affect the result, especially as far as the rise time is concerned. Indeed, a fit with a one-frame bin time (1.84\,s) tends to give a shorter rise $t_\mathrm{rise}$=(1.6--3.6)\,s, but clear residual structures 
appear, making any attempt to further constrain the rise-time inconclusive with the current \chandra~TE mode.\\

\subsection{Persistent and type-I burst spectral analysis}\label{sec:spectrum}

We extracted the spectra of the {\sl persistent} (non-burst) emission as well as of the type-I X-ray burst, hereafter {\sl burst-all}. 
Furthermore, we  divided the burst in three segments that we call hereafter {\sl rise} (about the first 4\,s), {\sl peak} (the next 13\,s) and {\sl tail} (the next 26\,s)\footnote{The final $\sim$10\,s of the burst have not been studied separately due to the extremely poor statistics.}. These times were chosen also taking into account  the more natural CCD-related read-out time frame, since the timed exposure (TE) mode configuration  is not designed to accurately sample fast variability. 

For each of the five parts ({\sl persistent, burst-all, rise, peak, tail}), we extracted
the first order dispersed spectra ($m=\pm1$ for HEG and MEG) and to 
increase the signal-to-noise ratio, we merged the two HEG ($m=\pm1$) and MEG ($m=\pm1$) spectra into one 
combined spectrum, for a total of five spectra (one per part)\footnote{The \chandra \zth-order~spectrum was not used in the spectral analysis as it severely suffers from pile-up, especially in the burst phase.}. 
Final binning, starting at 0.8\,keV, was chosen to have 
a signal to noise ratio higher than 5 and a minimum of 16 MEG channels per bin.  

Within the type-I X-ray burst, while we are confident that the {\sl burst-all}, {\sl peak} and {\sl tail} portions (54, 13 and 26\,s, respectively) include no time-tag shift due to the CCDs read-out,  there may potentially be a problem when integrating something varying as fast as 4\,s ({\sl rise}). Hence, though we have extracted its spectrum for visualization purposes, we have decided not to perform spectral fitting of the {\sl rise} bit.

\begin{figure}
\epsscale{1.2} \plotone{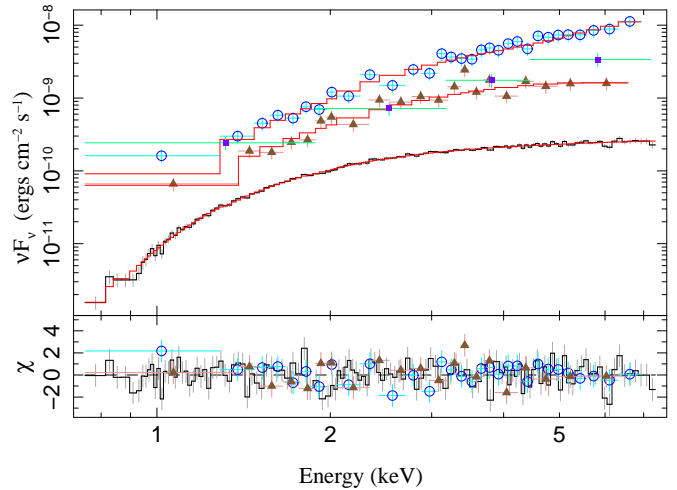}
\caption{Merged HEG ($m=\pm1$) and MEG ($m=\pm1$) spectra for the four extracted segments (see text): {\sl persistent} (black crosses, the dimmest one), {\sl peak} (blue circles, the brightest one), 
{\sl tail} (brown triangles) and {\sl rise} with widest energy binning (green).}
 \label{fig:fit}
\end{figure}

Figure~\ref{fig:fit} shows  four of the five spectra we obtain: {\sl persistent} (black crosses, the dimmest one), {\sl peak} (blue circles, the brightest one), 
{\sl tail} (brown triangles), and the shortest, 4\,s exposure, spectrum of the {\sl rise} with widest energy binning (green squares).
\begin{deluxetable*}{lccccccccr}
\setlength{\tabcolsep}{0.03in} 
\tabletypesize{\footnotesize}    
\tablewidth{0pt} 
\tablecaption{Fits to \mysou~Spectra: {\tt tbabs}*(\nthcomp+\bbodyrad)\label{tab:efits}}
\tablehead{
     \colhead{}
   & \colhead{$N_\mathrm{H}$\tablenotemark{(a)}}
   & \colhead{$kT_\mathrm{s}$}
   & \colhead{$\Gamma$} 
  & \colhead{$kT_\mathrm{e}$}
   & \colhead{$kT_\mathrm{bb}$}
   & \colhead{$R_\mathrm{bb}$\tablenotemark{(b)}} 
& \colhead{Average flux\tablenotemark{(c)}}
& \colhead{Average luminosity\tablenotemark{(d)}}
   & \colhead{$\chi^2/$DoF}
          \\                               
   & ($10^\mathrm{22}~\mathrm{cm^\mathrm{-2}}$) 
   & (keV)
   & 
   & (keV)
   & (keV)
   & (km)
  & ($\mathrm{10^{-10}~erg~cm^{-2}~s^{-1}}$)&
    (10$^{37}$ $\mathrm{erg~s^{-1}}$)     }
\startdata
   {\sl Persistent}
 & 1.02$\pm$0.05      		 % N_H
 & 0.55$\pm$0.03		% kTs
 & \errtwo{1.79}{0.07}{0.06}     % Powerlaw Slope
 & [50]                      % kTe
 & -          % kTbb
 & -          % Rbb
 & 3.17    % Chandra Flux
 & 0.17    % Chandra L
 & 178.2/178                     % Chi^2/DoF
\\
%\noalign{\vspace*{0.2mm}}
%   {\sl Raise} 
% & \nodata                     % N_H
% & \nodata                     % kTs
% & \nodata % Powerlaw Slope
% & \nodata % kTbb
% & \nodata % Chandra flux
% & \nodata % ISGRI flux
% & \nodata                      % Chi^2/DoF
%\\
\noalign{\vspace*{0.4mm}}
   {\sl Burst peak} (13\,s)
 & [1.02]     			% N_H
 & [0.55]		% kTs
 & [1.79]                     % Powerlaw Slope
 & [50]                      % kTe
 &  \errtwo{2.5}{0.8}{0.4}                  % kTbb
 & 5$\pm$1      % Rbb
 & 110   % Chandra Flux
 & 5.93    % Chandra L
 & \nodata                    % Chi^2/DoF
\\
\noalign{\vspace*{0.4mm}}
    {\sl Burst tail} (26\,s)
 & [1.02]    				% N_H
 &  [0.55]			% kTs
 &   [1.79]                       % Powerlaw Slope
 & [50]                      % kTe
 & \errtwo{1.3}{0.2}{0.1}                  % kTbb
 & 5.3$\pm$0.8         % Rbb
 & 18.9    % Chandra Flux
 & 1.02  % Chandra L
 & \nodata                   % Chi^2/DoF
\\
\noalign{\vspace*{0.4mm}}
   {\sl Burst-all} (54\,s)
 &  [1.02]   			% N_H
 &  [0.55]		% kTs
 &  [1.79]                     % Powerlaw Slope
 &  [50]                      % kTe
 &  1.6$\pm$0.1                  % kTbb
 & 5.2$\pm$0.5     % Rbb
 &  28.6   % Chandra Flux
 & 1.6    % Chandra L
 & \nodata                    % Chi^2/DoF
\\

  \tablecomments{Errors bars are 90\% confidence level for one 
     parameter. The input seed photons to the \nthcomp~model are blackbody in shape. } 

\tablenotetext{(a)} {In the fit we have used an improved model for the absorption of X-rays in the interstellar medium by \cite{wilms00}.} 
\tablenotetext{(b)} {Assuming a distance of 6.9\,kpc \citep{altamirano10}.} 
\tablenotetext{(c)}{Absorbed 0.5--8\,keV flux. } 
\tablenotetext{(d)}{Absorbed 0.5--8\,keV luminosity, assuming a distance of 6.9\,kpc.}

\end{deluxetable*}

The fit of the {\sl persistent} spectrum with a single non-Comptonised component, be it blackbody or disk blackbody, was very poor, with clear structured residuals. Hence we used a thermal Comptonization model  \citep[\nthcomp in \xspec terminology,][]{zdziarski96,zycki99}, where soft seed photons of temperature $kT_\mathrm{s}$ are up-scattered by a thermal population of electrons at a temperature of $kT_\mathrm{e}$.
Since the hot electrons up-scatter the seed photons, there are few photons remaining at energies below the typical seed photon energies, making  it significantly different from a power-law below this energy. However, the spectrum can be parameterized by an asymptotic power-law index ($\Gamma$) that is also a parameter in the model, together with $kT_\mathrm{s}$, $kT_\mathrm{e}$ and the model normalization. 

 While the low energy roll-over of the spectrum, related to $kT_\mathrm{s}$, can be well appreciated in the \chandra energy range, the higher energy one, related to $kT_\mathrm{e}$, is known to be out of the current range \citep[e.g.,][]{papitto10}. Since our spectral fits are not sensitive to its value, we choose to freeze it to $kT_\mathrm{e}$=50\,keV in the spectral fit. In the model, the seed photons can be blackbody or disk blackbody but since we cannot discriminate between the two with the current data, we choose to report only the blackbody shape case.
Figure~\ref{fig:fit} shows the best fit we obtained with the absorbed thermal Comptonization model, while Table~1 shows the obtained parameters. In this case, $kT_\mathrm{bb}$ and $R_\mathrm{bb}$ of Table~1 are not applicable, since no additional thermal component is required in the {\sl persistent} spectrum (see section \ref{sec:persist}).

The fits of the {\sl burst-all}, {\sl peak} and {\sl tail} segments were obtained adding to the above {\sl persistent} model, fixed and considered as the continuum, an additional blackbody component (\bbodyrad~in \xspec terminology). The obtained best fit, temperatures and radii \citep[assuming a distance of 6.9\,kpc,][]{altamirano10} can also be seen in Table~1. 

We note that in all the spectral fitting of Table~1 the additional function {\sl simple\_gpile2} was applied to the spectra in order to correct for pile-up distortions, as explained in \cite{nowak08} and \cite{hanke09}.

\section{Discussion}\label{sec:discussion}

\subsection{The persistent emission}\label{sec:persist}
The broad-band spectra of AMXPs can normally be described as the composition of an accretion disk emission (peaking below $\sim$2\,keV), a blackbody originating from the hot spot  and a hard X-ray emission originated by thermal Comptonisation contributing to the whole broad-band 1-200\,keV emission \citep[e.g., see Figure~1 in ][and references therein]{ibragimov11}. 
As an example of the temperatures involved, the joint \xmm-\rxte~spectrum of \mysou~as observed by \cite{papitto10} could be modeled by 
those three components that were interpreted, from the softest to the hardest, as a multicolored disk 
emission ($kT_\mathrm{in}=0.36\pm$0.2\,keV), thermal emission from the neutron star surface 
($kT_\mathrm{BB}$=\errtwo{0.64}{0.01}{0.02}\,keV) and thermal Comptonisation emission of hotter seed 
photons ($kT_\mathrm{seed}$=\errtwo{1.37}{0.01}{0.02}\,keV) by a hot plasma of electrons 
($kT_\mathrm{e}$=\errtwo{51}{6}{4}\,keV, $\tau$=\errtwo{1.34}{0.03}{0.06})\footnote{The authors 
obtain the well constrained Comptonization parameters using simultaneous 
\xmm-\rxte~data.}. 

In our 20\,ksec \chandra observation, due to lower statistics, the presence of more than one 
thermal emission component is not required by the data. Indeed, we obtain  a single soft population of 
$kT_\mathrm{s}$=0.55$\pm$0.03\,keV as seed photons for the Comptonisation, most likely the non disentangled combination of  accretion disk and neutron star surface/halo thermal populations, with
no additional thermal component required. A 90\% upper limit on the normalization of an additional thermal component, e.g., \bbodyrad~model (with temperature fixed to the Comptonization component of Table 1) gives  an absorbed 0.5--8\,keV flux of $1.7\times\mathrm{10^{-11}~erg~cm^{-2}~s^{-1}}$, to be compared with our persistent emission of $3.17\times\mathrm{10^{-10}~erg~cm^{-2}~s^{-1}}$. This is consistent with what \cite{papitto10} obtain using a 70\,ksec \xmm~observation\footnote{We refer to Table~2 in \cite{papitto10}, model A, the closest to the one we have here.}, i.e., a total thermal absorbed 0.5--8\,keV flux 
%(in their case \diskbb$+$\bbodyrad~in addition to \nthcomp) 
of about $1.3\times\mathrm{10^{-11}~erg~cm^{-2}~s^{-1}}$, besides the Comptonised \nthcomp component. Furthermore, adding a \diskline~component and  freezing its values (except normalization) to the best fit from \citet[][Table 2, \xmm~data only]{papitto10}, we obtain a  90\% upper limit on the equivalent width of 50\,eV, consistent with  that measured by \xmm, 43.9$\pm$0.06\,eV.

The 0.5--8\,keV spectrum of \mysou~is energetically dominated by a power-law which is equivalent to a broader Comptonised emission on a limited bandwidth between $kT_\mathrm{s}$ and $kT_\mathrm{e}$. In the case of \nthcomp, the code provides, as the best-fit parameters, the seed photon temperature $kT_\mathrm{s}$, the electron temperature   $kT_\mathrm{e}$ and the power-law spectral index $\Gamma$.

 Since \chandra~limits us to studying the 0.8--8\,keV range, the temperature of the Comptonising plasma was fixed to $kT_\mathrm{e}$=50\,keV, while the photon index of the power-law, that dominates the spectrum, was obtained by the fit as $\Gamma$= \errtwo{1.79}{0.07}{0.05}. Since no high energy cutoff appears in our \chandra~spectrum of \mysou, the choice of fixing the $kT_\mathrm{e}$ parameter to 50\,keV does not affect our results. An assumed temperature of, e.g., 100\,keV yields  comparable results and $\Gamma$ value.

Once $kT_\mathrm{e}$ and $\Gamma$ are provided, it is possible to infer the Thomson optical 
depth $\tau$ through the relation:

\begin{equation}
\label{eq:gamma}
\Gamma=\left[ \frac{9}{4}+\frac{1}{\left(\frac{kT_e}{m_ec^2}\right)\tau\left(1+\frac{\tau}{3}\right)}\right]^{1/2}-\frac{1}{2}
\end{equation} (see, e.g., \citealt{lightman87}).
We obtain values of optical depth $\tau$=(1--2) for 
$kT_\mathrm{e}$=(50--100\,keV), similarly to what obtained by \citet{papitto10}.

 Using the {\sl persistent} model obtained in Table~1, we obtain 
 $L_\mathrm{0.5-8\,keV}$=1.7$\times$10$^{36}$ $\mathrm{erg~s^{-1}}$ (at 6.9\,kpc) and an extrapolated unabsorbed  $L_\mathrm{2-200\,keV}$=7$\times$10$^{36}$ $\mathrm{erg~s^{-1}}$. This is consistent with Figure~4 in \cite{altamirano10}, where a persistent $L_\mathrm{2-200\,keV}\sim$(7--8)$\times$10$^{36}$ $\mathrm{erg~s^{-1}}$ is expected from the source at 6.9\,kpc in the time lasting between the fifth and sixth \rxte~burst, when the \chandra~observation occurred.

A more detailed analysis of the persistent spectrum is not justified by the data. No low energy features are visible and since the distribution of the residuals does not show any systematic trend, we believe that the best fit model we obtain (Fig.\ref{fig:fit} and Table~1) is the simplest and most coherent description of the data, with results compatible with what found in literature, albeit subject to uncertainties due to model extrapolations and comparison of different mission calibrations.  

\subsection{The type-I burst}\label{sec:discussionburst}

During our \chandra~observation, a type-I X-ray burst  was observed (Fig.\ref{fig:burst}). The burst, fit with a FRED function resulted in a rise time $t_\mathrm{rise}$=(3.6--6.3)\,s (but see section \ref{sec:theburst}) and decay $t_\mathrm{decay}$=(14$\pm$1)\,s. 
This is \lq\lq slow\rq\rq~if compared to the ten type-I X-ray bursts of \mysou~observed in the \rxte/PCA data \citep{altamirano10}, where all the bursts reached their maxima within 1.2\,s  and with decay times in the range of 5--8\,s.  Although we cannot exclude that this burst is slightly longer than the \rxte reported ones, we note that a more accurate  comparison is hampered by the limitations of the \chandra~timed exposure (TE) mode, where each chip is exposed for approximately 2\,s, to be compared with the 0.1\,s time resolution of the \rxte~data of \cite{altamirano10} (see section \ref{sec:theburst}).

The low statistics obtained in the burst prevented us from doing an accurate phase resolved spectroscopy as done in the case of, e.g., the brightest \rxte~burst \citep{altamirano10,falanga11} or in the \xmm~ones \citep{papitto10}. In our case it was only possible to split the burst in three segments ({\sl rise, peak, tail}), fitting only the latter two because of instrumental limitations (see section \ref{sec:spectrum}). 
As shown in Table~1, the burst emission could be well fit by a single blackbody with the temperature decaying from $kT_\mathrm{bb}$=\errtwo{2.5}{0.8}{0.4}\,keV to $kT_\mathrm{bb}$=\errtwo{1.3}{0.2}{0.1}\,keV,   likely indicating the cooling of the neutron star surface after the burst ignition. The related emitting area has in both cases a radius comparable with $\sim$5\,km that is consistent with that found by, e.g., \cite{falanga11} and \cite{papitto10} for the other bursts from \mysou. We note however that direct comparison with the other X-ray bursts is to be taken with caution, because unlike in the other cases, where a detailed phase resolved analysis was possible, in our study the obtained quantities of Table~1 are averaged on large portions of the burst (e.g., 13\,s for the {\sl peak}). 

Nevertheless, a comparison of the overall properties of our type-I burst with the ones previously reported using \rxte~and \integral~can be attempted.
 Using the {\sl peak} model obtained in Table~1 (with $N_\mathrm{H}$ and the \nthcomp normalisation set to 0), we obtain a {\sl peak} unabsorbed luminosity
$L_\mathrm{0.1-40\,keV}\sim$1.5$\times$10$^{38}$ $\mathrm{erg~s^{-1}}$  for the source at 6.9\,kpc. This is a 13\,s average value and to compare it  with the non averaged \rxte~ones, we should estimate our \lq\lq real peak\rq\rq~value, obtaining it from the FRED function that best fits our burst profile. This results in $L_\mathrm{0.1-40\,keV}^\mathrm{peak}\sim$2.3$\times$10$^{38}$ $\mathrm{erg~s^{-1}}$, to be compared to 
 Figure~4, middle panel, in \cite{altamirano10}, where a bolometric peak $L\sim$(2.5--3.5)$\times$10$^{38}$ $\mathrm{erg~s^{-1}}$ is expected from the source at 6.9\,kpc, during the \chandra observation, between the fifth and sixth \rxte~burst.

To obtain an estimate of the overall type-I burst total energy release ($E_\mathrm{0.1-40\,keV}^\mathrm{burst}$) and fluence ($f_\mathrm{0.1-40\,keV}^\mathrm{burst}$), to be compared with the results of the \rxte~bursts by \cite{altamirano10} and of the \rxte-\integral~ones by \cite{falanga11}, we consider the {\sl burst-all} spectrum of Table~1. Setting $N_\mathrm{H}$ and the \nthcomp normalisation to 0, we obtain a bolometric unabsorbed luminosity of $L_\mathrm{0.1-40\,keV}^\mathrm{burst}\sim$2$\times$10$^{37}\mathrm{erg~s^{-1}}$ that results in a total energy release of $E_\mathrm{0.1-40\,keV}^\mathrm{burst}\sim$1.1$\times$10$^{39}$\,erg in 54\,s. This is slightly lower than the range obtained by \cite{altamirano10}, (2.5--3)$\times$10$^{39}$\,erg. Similarly, a fluence of  $f_\mathrm{0.1-40\,keV}^\mathrm{burst}$=2$\times$10$^{-7}\mathrm{erg~cm^{-2}}$ is obtained, to be compared to (3.2--4.2)$\times$10$^{-7}\mathrm{erg~cm^{-2}}$ of \cite{falanga11}. In both cases we are dimmer than the previously reported X-ray bursts. Indeed, our burst could be intrinsically dimmer, however considering that we are subject to uncertainties in the model extrapolations beyond the \chandra~energy domain, as well as to uncertainties in the mission cross-calibrations, it is reasonable to conclude that we have no strong evidence for the type-I X-ray burst observed by \chandra~from \mysou~to be inconsistent with the previously reported bursts. Furthermore, the time averaging issue reported above and in Section~\ref{sec:theburst} due to the timed exposure (TE) mode of the observation may be the dominant source of discrepancy. Indeed a limited number of type-I X-ray bursts have been studied up to now with \chandra~grating;  most observations were done in continuous clocking mode (CC), for which the ACIS-S CCDs are read out continuously, providing a $\sim$3\,msec timing, at the expense of one dimension of spatial resolution. An example is, e.g., the study of radius-expansion burst spectra from 4U~1728$-$34 \citep{galloway10}. Out of the 25 bursts detected, time-resolved spectroscopy of the summed signal from the four brightest bursts (with summed bolometric peak flux of about $8\times\mathrm{10^{-8}~erg~cm^{-2}~s^{-1}}$) was carried out. A clear photospheric radius expansion in these bursts could be seen, well sampled over seven time bins on a total of 12\,sec of burst duration. Stacking data-sets from several \chandra~burst intervals for a more detailed spectral study and evolution has also been the approach of  \cite{thompson05} for GS~1826$-$238. Similarly to our case, TE mode had been used and indeed time bins of about 10\,sec (minimum) were extracted for a spectral study on the six averaged type-I bursts detected from the source (lasting about 150\,sec). \\
The \chandra~observation of \mysou~presented here was done in TE mode and a single unexpected 54\,sec type-I X-ray burst was detected. In this work, we have carefully described our approach on analyzing the lightcurve and spectra, as the burst is
evolving, also showing what can be done within the confines of such an observation.

\acknowledgments 
We thank the anonymous Referee for useful comments that greatly improved the quality of the paper.
We thank the \chandra team for their rapid response in
scheduling and delivering the observation. 
This research has made use of the \integral~sources page maintained by J. Rodriguez \& A. Bodaghee (http://irfu.cea.fr/Sap/IGR-Sources/). AP and PU acknowledge financial contribution from the ASI-INAF agreements I/009/10/0 and I/033/10/0. MDS  acknowledges financial contribution from PRIN-INAF 2009 (PI: L. Sidoli) and from the ASI-INAF agreement I/009/10/0. AP acknowledges John Houck for his precious support in the \chandra/ISIS software installation phase.

\newpage
\bibliographystyle{apj}
\bibliography{biblio}

\begin{thebibliography}{37}
\expandafter\ifx\csname natexlab\endcsname\relax\def\natexlab#1{#1}\fi

\bibitem[{{Altamirano} {et~al.}(2008){Altamirano}, {Casella}, {Patruno},
  {Wijnands}, \& {van der Klis}}]{altamirano08}
{Altamirano}, D., {Casella}, P., {Patruno}, A., {Wijnands}, R., \& {van der
  Klis}, M. 2008, \apjl, 674, L45

\bibitem[{{Altamirano} {et~al.}(2010){Altamirano}, {Watts}, {Linares},
  {Markwardt}, {Strohmayer}, \& {Patruno}}]{altamirano10}
{Altamirano}, D., {Watts}, A., {Linares}, M., {Markwardt}, C.~B., {Strohmayer},
  T., \& {Patruno}, A. 2010, \mnras, 409, 1136

\bibitem[{{Baldovin} {et~al.}(2009){Baldovin}, {Kuulkers}, {Ferrigno}, {Bozzo},
  {Chenevez}, {Brandt}, {Beckmann}, {Bird}, {Domingo}, {Ebisawa}, {Jonker},
  {Kretschmar}, {Markwardt}, {Oosterbroek}, {Paizis}, {Risquez},
  {Sanchez-Fernandez}, {Shaw}, \& {Wijnands}}]{baldovin09}
{Baldovin}, C., {et~al.} 2009, The Astronomer's Telegram, 2196

\bibitem[{{Bozzo} {et~al.}(2009){Bozzo}, {Ferrigno}, {Kuulkers}, {Falanga},
  {Chenevez}, {Brandt}, {Beckmann}, {Bird}, {Domingo}, {Ebisawa}, {Jonker},
  {Kretschmar}, {Markwardt}, {Oosterbroek}, {Paizis}, {Risquez},
  {Sanchez-Fernandez}, {Shaw}, \& {Wijnands}}]{bozzo09}
{Bozzo}, E., {et~al.} 2009, The Astronomer's Telegram, 2198

\bibitem[{{Canizares} {et~al.}(2000){Canizares}, {Huenemoerder}, {Davis},
  {Dewey}, {Flanagan}, {Houck}, {Markert}, {Marshall}, {Schattenburg},
  {Schulz}, {Wise}, {Drake}, \& {Brickhouse}}]{canizares00}
{Canizares}, C.~R., {et~al.} 2000, \apjl, 539, L41

\bibitem[{{Casella} {et~al.}(2008){Casella}, {Altamirano}, {Patruno},
  {Wijnands}, \& {van der Klis}}]{casella08}
{Casella}, P., {Altamirano}, D., {Patruno}, A., {Wijnands}, R., \& {van der
  Klis}, M. 2008, \apjl, 674, L41

\bibitem[{{Falanga} {et~al.}(2011){Falanga}, {Kuiper}, {Poutanen}, {Galloway},
  {Bonning}, {Bozzo}, {Goldwurm}, {Hermsen}, \& {Stella}}]{falanga11}
{Falanga}, M., {et~al.} 2011, \aap, 529, A68

\bibitem[{{Galloway} {et~al.}(2010){Galloway}, {Yao}, {Marshall}, {Misanovic},
  \& {Weinberg}}]{galloway10}
{Galloway}, D.~K., {Yao}, Y., {Marshall}, H., {Misanovic}, Z., \& {Weinberg},
  N. 2010, \apj, 724, 417

\bibitem[{{Hanke} {et~al.}(2009){Hanke}, {Wilms}, {Nowak}, {Pottschmidt},
  {Schulz}, \& {Lee}}]{hanke09}
{Hanke}, M., {Wilms}, J., {Nowak}, M.~A., {Pottschmidt}, K., {Schulz}, N.~S.,
  \& {Lee}, J.~C. 2009, \apj, 690, 330

\bibitem[{{Houck}(2002)}]{houck02}
{Houck}, J.~C. 2002, in High Resolution X-ray Spectroscopy with XMM-Newton and
  Chandra

\bibitem[{{Huenemoerder} {et~al.}(2011){Huenemoerder}, {Mitschang}, {Dewey},
  {Nowak}, {Schulz}, {Nichols}, {Davis}, {Houck}, {Marshall}, {Noble},
  {Morgan}, \& {Canizares}}]{huenemoerder11}
{Huenemoerder}, D.~P., {et~al.} 2011, \aj, 141, 129

\bibitem[{{Ibragimov} {et~al.}(2011){Ibragimov}, {Kajava}, \&
  {Poutanen}}]{ibragimov11}
{Ibragimov}, A., {Kajava}, J.~J.~E., \& {Poutanen}, J. 2011, \mnras, 415, 1864

\bibitem[{{Kaaret} {et~al.}(2006){Kaaret}, {Morgan}, {Vanderspek}, \&
  {Tomsick}}]{kaaret06}
{Kaaret}, P., {Morgan}, E.~H., {Vanderspek}, R., \& {Tomsick}, J.~A. 2006,
  \apj, 638, 963

\bibitem[{{Kalamkar} {et~al.}(2011){Kalamkar}, {Altamirano}, \& {van der
  Klis}}]{kalamkar11}
{Kalamkar}, M., {Altamirano}, D., \& {van der Klis}, M. 2011, \apj, 729, 9

\bibitem[{{Kuulkers} {et~al.}(2007){Kuulkers}, {Shaw}, {Paizis}, {Chenevez},
  {Brandt}, {Courvoisier}, {Domingo}, {Ebisawa}, {Kretschmar}, {Markwardt},
  {Mowlavi}, {Oosterbroek}, {Orr}, {R{\'{\i}}squez}, {Sanchez-Fernandez}, \&
  {Wijnands}}]{kuulkers07}
{Kuulkers}, E., {et~al.} 2007, \aap, 466, 595

\bibitem[{{Lightman} \& {Zdziarski}(1987)}]{lightman87}
{Lightman}, A.~P., \& {Zdziarski}, A.~A. 1987, \apj, 319, 643

\bibitem[{{Liu} {et~al.}(2007){Liu}, {van Paradijs}, \& {van den
  Heuvel}}]{liu07}
{Liu}, Q.~Z., {van Paradijs}, J., \& {van den Heuvel}, E.~P.~J. 2007, \aap,
  469, 807

\bibitem[{{Markwardt} {et~al.}(2009){Markwardt}, {Altamirano}, {Strohmayer}, \&
  {Swank}}]{markwardt09}
{Markwardt}, C.~B., {Altamirano}, D., {Strohmayer}, T.~E., \& {Swank}, J.~H.
  2009, The Astronomer's Telegram, 2237

\bibitem[{{Miller-Jones} {et~al.}(2009){Miller-Jones}, {Russell}, \&
  {Migliari}}]{miller09}
{Miller-Jones}, J.~C.~A., {Russell}, D.~M., \& {Migliari}, S. 2009, The
  Astronomer's Telegram, 2232

\bibitem[{{Nowak} {et~al.}(2008){Nowak}, {Juett}, {Homan}, {Yao}, {Wilms},
  {Schulz}, \& {Canizares}}]{nowak08}
{Nowak}, M.~A., {Juett}, A., {Homan}, J., {Yao}, Y., {Wilms}, J., {Schulz},
  N.~S., \& {Canizares}, C.~R. 2008, \apj, 689, 1199

\bibitem[{{Nowak} {et~al.}(2009){Nowak}, {Paizis}, {Wilms}, {Rodriguez},
  {Chaty}, {Ebisawa}, {Del Santo}, {Farinelli}, {Ubertini}, \&
  {Courvoisier}}]{nowak09}
{Nowak}, M.~A., {et~al.} 2009, The Astronomer's Telegram, 2215

\bibitem[{{Papitto} {et~al.}(2010){Papitto}, {Riggio}, {di Salvo}, {Burderi},
  {D'A{\`i}}, {Iaria}, {Bozzo}, \& {Menna}}]{papitto10}
{Papitto}, A., {Riggio}, A., {di Salvo}, T., {Burderi}, L., {D'A{\`i}}, A.,
  {Iaria}, R., {Bozzo}, E., \& {Menna}, M.~T. 2010, \mnras, 407, 2575

\bibitem[{{Papitto} {et~al.}(2011){Papitto}, {Bozzo}, {Ferrigno}, {Belloni},
  {Burderi}, {di Salvo}, {Riggio}, {D'A{\`i}}, \& {Iaria}}]{papitto11}
{Papitto}, A., {et~al.} 2011, \aap, 535, L4

\bibitem[{{Patruno}(2010{\natexlab{a}})}]{patruno10b}
{Patruno}, A. 2010{\natexlab{a}}, ArXiv 1007.1108

\bibitem[{{Patruno}(2010{\natexlab{b}})}]{patruno10}
---. 2010{\natexlab{b}}, \apj, 722, 909

\bibitem[{{Primini} {et~al.}(2011){Primini}, {Houck}, {Davis}, {Nowak},
  {Evans}, {Glotfelty}, {Anderson}, {Bonaventura}, {Chen}, {Doe}, {Evans},
  {Fabbiano}, {Galle}, {Gibbs}, {Grier}, {Hain}, {Hall}, {Harbo}, {(Helen He},
  {Karovska}, {Kashyap}, {Lauer}, {McCollough}, {McDowell}, {Miller},
  {Mitschang}, {Morgan}, {Mossman}, {Nichols}, {Plummer}, {Refsdal}, {Rots},
  {Siemiginowska}, {Sundheim}, {Tibbetts}, {Van Stone}, {Winkelman}, \&
  {Zografou}}]{primini11}
{Primini}, F.~A., {et~al.} 2011, \apjs, 194, 37

\bibitem[{{Riggio} {et~al.}(2011){Riggio}, {Papitto}, {Burderi}, {di Salvo},
  {Bachetti}, {Iaria}, {D'A{\`i}}, \& {Menna}}]{riggio11}
{Riggio}, A., {Papitto}, A., {Burderi}, L., {di Salvo}, T., {Bachetti}, M.,
  {Iaria}, R., {D'A{\`i}}, A., \& {Menna}, M.~T. 2011, \aap, 526, A95

\bibitem[{{Rodriguez} {et~al.}(2008){Rodriguez}, {Shaw}, {Hannikainen},
  {Belloni}, {Corbel}, {Cadolle Bel}, {Chenevez}, {Prat}, {Kretschmar},
  {Lehto}, {Mirabel}, {Paizis}, {Pooley}, {Tagger}, {Varni{\`e}re}, {Cabanac},
  \& {Vilhu}}]{rodriguez08}
{Rodriguez}, J., {et~al.} 2008, \apj, 675, 1449

\bibitem[{{Strohmayer} \& {Bildsten}(2006)}]{strohmayer06}
{Strohmayer}, T., \& {Bildsten}, L. 2006, {New views of thermonuclear bursts},
  ed. {Lewin, W.~H.~G.~\& van der Klis, M.}, 113--156

\bibitem[{{Tauris} \& {van den Heuvel}(2006)}]{tauris06}
{Tauris}, T.~M., \& {van den Heuvel}, E.~P.~J. 2006, {Formation and evolution
  of compact stellar X-ray sources}, ed. {Lewin, W.~H.~G.~\& van der Klis, M.},
  623--665

\bibitem[{{Thompson} {et~al.}(2005){Thompson}, {Rothschild}, {Tomsick}, \&
  {Marshall}}]{thompson05}
{Thompson}, T.~W.~J., {Rothschild}, R.~E., {Tomsick}, J.~A., \& {Marshall},
  H.~L. 2005, \apj, 634, 1261

\bibitem[{{Torres} {et~al.}(2009{\natexlab{a}}){Torres}, {Jonker}, {Steeghs},
  {Damjanov}, {Caris}, \& {Glazebrook}}]{torres11b}
{Torres}, M.~A.~P., {Jonker}, P.~G., {Steeghs}, D., {Damjanov}, I., {Caris},
  E., \& {Glazebrook}, K. 2009{\natexlab{a}}, The Astronomer's Telegram, 2233

\bibitem[{{Torres} {et~al.}(2009{\natexlab{b}}){Torres}, {Jonker}, {Steeghs},
  {Simon}, \& {Gutowski}}]{torres11a}
{Torres}, M.~A.~P., {Jonker}, P.~G., {Steeghs}, D., {Simon}, J.~D., \&
  {Gutowski}, G. 2009{\natexlab{b}}, The Astronomer's Telegram, 2216

\bibitem[{{Watts} {et~al.}(2009){Watts}, {Altamirano}, {Markwardt}, \&
  {.}}]{watts09}
{Watts}, A.~L., {Altamirano}, D., {Markwardt}, C.~B., \& {.}, T.~E.~S. 2009,
  The Astronomer's Telegram, 2199

\bibitem[{{Wilms} {et~al.}(2000){Wilms}, {Allen}, \& {McCray}}]{wilms00}
{Wilms}, J., {Allen}, A., \& {McCray}, R. 2000, \apj, 542, 914

\bibitem[{{Zdziarski} {et~al.}(1996){Zdziarski}, {Johnson}, \&
  {Magdziarz}}]{zdziarski96}
{Zdziarski}, A.~A., {Johnson}, W.~N., \& {Magdziarz}, P. 1996, \mnras, 283, 193

\bibitem[{{{\.Z}ycki} {et~al.}(1999){{\.Z}ycki}, {Done}, \& {Smith}}]{zycki99}
{{\.Z}ycki}, P.~T., {Done}, C., \& {Smith}, D.~A. 1999, \mnras, 309, 561

\end{thebibliography}

\end{document}